\documentclass[11pt]{article}
\usepackage[margin=1in]{geometry}	% use reasonable margins, not the stupid latex defaults
\usepackage{setspace}				% global line spacing except for footnotes and floats (figs, tables)
\usepackage[natbibapa]{apacite} % apa style citations with natbib citation options

\usepackage[latin1]{inputenc}
\usepackage{amsmath,amsfonts, amssymb}   % ams math, fonts and symbols
\usepackage{verbatim} 
\usepackage[english]{babel}
\usepackage[T1]{fontenc}
\usepackage{bbm,bm}

\usepackage{lmodern}
\usepackage{blindtext}

\usepackage{float}

\usepackage{graphicx, floatflt} %, subfig}
\usepackage[labelformat=simple]{subcaption}

\DeclareCaptionLabelFormat{subcaptionlabel}{\normalfont(\textbf{#2}\normalfont)}
\usepackage{color}				% Colored text. Does it make sense to load  both xcolor and color?
\usepackage[normalem]{ulem}	% support text strikethrough
\usepackage{cancel}				% support equation strikethroug
\usepackage[export]{adjustbox}

\usepackage{enumitem}% separation of lines within \begin{itemize} ... \end{itemize}

\usepackage{caption}
\usepackage{multirow}
\usepackage{booktabs}
\usepackage{adjustbox}
\usepackage{esint}
\usepackage[title]{appendix}
\providecommand{\keywords}[1]
{
  \small	
  \textbf{ \\ Keywords:} #1
}

\graphicspath{{Figures/}}

\newcommand{\one}{\mathbbm{1} }

\DeclareMathOperator*{\argmin}{arg\,min} 	% thin space between arg and min, limits placed underneath

\usepackage{authblk}% supports both author block or inline with footnote markers

\title{\bf An Empirical Implementation of the Shadow Riskless Rate}
\author[1]{Davide Lauria}
\author[2]{Jiho Park}
\author[3]{Yuan Hu}
\author[4,*]{W. Brent Lindquist}
\author[4]{Svetlozar T. Rachev}
\author[5]{Frank J. Fabozzi}

\affil[1]{\small Department of Economics, Statistics \& Finance, University of Calabria, via Pietro Bucci, Arcavacata di Rende, Calabria 87036, Italy; davide.lauria@unical.it}
\affil[2]{\small Market Risk Analytics, Citigroup, 6460 Las Colinas Blvd, Irving, TX 75039, USA; jihopark19@gmail.com}
\affil[3]{\small Independent Researcher, 1620 E. Jefferson St. 312, Rockville, MD 20852, USA; yuanhu0326@gmail.com}
\affil[4]{\small Department of Mathematics \& Statistics, Texas Tech University, Lubbock, TX 79409-1042, USA;
brent.lindquist@ttu.edu; zari.rachev@ttu.edu}
\affil[5]{\small Carey Business School, Johns Hopkins University, Baltimore, MD 21202, USA; fabozzi321@aol.com}

\affil[*]{Corresponding author, brent.lindquist@ttu.edu}

\begin{document}
\maketitle

\abstract{
We address the problem of asset pricing in a market where there is no risky asset.
Previous work developed a theoretical model for a shadow riskless rate (SRR) for such a market
in terms of the drift component of the state-price deflator for that asset universe.
Assuming asset prices are modeled by correlated geometric Brownian motion,
in this work we develop a computational approach  to estimate  the SRR from empirical datasets.
The approach employs: principal component analysis to model the effects of the individual Brownian motions;
singular value decomposition to capture the abrupt changes in condition number of the linear system whose
solution provides the SRR values;
and a regularization to control the rate of change of the condition number.
Among other uses (e.g., for option pricing, developing a term structure of interest rate),
the SRR can be employed as an investment discriminator between asset classes.
We apply the computational procedure to markets consisting of groups of stocks, varying asset type and number.
The theoretical and computational analysis provides not only the drift, but also the total volatility of the state-price deflator.
We investigate the time trajectory of these two descriptive components of the state-price deflator for the empirical datasets.
}

\keywords{riskless rate; safe assets; geometric Brownian motion; state-price deflator; principal component analysis}

\section{Introduction}\label{sec:intro} 
One definition of a riskless (safe) asset is ``an asset that is (almost always) valued at face value without expensive and prolonged analysis'' \citep{Gorton_2017}.
A safe asset changes hands with ``no questions asked'' \citep{Holmstrom_2015}.
Riskless assets play a crucial role in financial markets, both in practice and in financial theory. 
Safe assets are used  by  financial entities to satisfy regulatory requirements, as pricing benchmarks,
as collateral in financial transactions, and in the development of the theory of asset and derivative pricing.
\cite{Gorton_2012} have shown that, while the total volume of US financial assets has increased 250\% since 1952,
 the percentage of safe debt in the US economy has remained relatively stable at approximated 32\%.
This stability in the percentage of safe debt obscures the fact that the composition of safe assets has changed,
from consisting primarily of government debt (US Treasury securities) and cash (demand deposits)
to an ever-increasing reliance on innovative financial instruments (e.g., money market mutual fund shares, commercial paper,
repurchase agreements, and securitized debt such as asset-backed and mortgaged-backed securities)
developed by the shadow banking system. 
The growth in the volume of financial assets, outpacing the availability of traditional safe assets,
reflects the fact that shadow institutions are not subject to the same regulations as depository banks,
do not need to keep the same level of financial reserves relative to market exposure,
and can maintain higher levels of financial leverage.
 
This vulnerability became apparent during the 2008 global financial crisis.
A consequence of subsequent regulations,
such as the US Dodd-Frank Act and the BCBS Basel III Accord,
has been a reduction in the availability of safe assets.
However, a number of studies
\citep{Eggertsson_2012, Gourinchas_2012, Caballero_2013, Aoki_2014, Caballero_2020, Caballero_2021,Gorton_2022,Bocola_2023,Caramp_2024}
have focused on the
challenges that an insufficiency of safe assets would have on the stability of the financial system and on monetary policy.
The work of these researchers suggests that the imposition of a liquidity coverage ratio that requires banks to
back their short-term debt dollar-for-dollar with treasury securities risks both lowering interest rates and making panics more frequent. 
\cite{Caballero_2018} describe how a self-reinforcing demand for safe assets which are in short supply might turn a
lengthy recession into outright stagnation.
\cite{Altinoglu_2023} introduced a theory of safe asset creation to study the interaction between systemic risk (including financial leveraging)
and aggregate demand (the restriction of the safe asset supply through external constraints) .

As early as 1972, economists considered equilibrium markets having no riskless assets. 
For example, \cite{Black_1972} showed that the capital asset pricing model is still valid even without a riskless asset.
Recognizing that the continued development of non-traditional safe assets is a
``train in motion that probably can't be stopped'',
it is necessary to replace qualitative definitions of a safe asset with a strong theoretical (and consequent empirical) construct.
In response to this,
\cite{Rachev_2017} developed a riskless rate and a corresponding riskless asset for a market consisting solely of $N$ risky assets.
Theoretically, this riskless rate is (the negative of) the drift term of the state-price deflator for that market.
They first considered the case where the price dynamics of the $N$ assets were correlated geometric
Brownian motions (GBMs) driven by $N-1$  Brownian motions.
In addition to this continuous diffusion case,
they also considered the cases in which the price dynamics of the assets were determined by: correlated
jump-diffusions; diffusions with stochastic volatilities; and geometric fractional  Brownian and Rosenblatt processes.
For each choice of price dynamics, they derived the Black--Scholes--Merton equations to price a European contingent claim,
constructing a tradeable, perpetual derivative which serves as a proxy riskless asset
whose price evolves according to this riskless rate.
The resultant market consisting of the $N$ risky assets and this perpetual derivative is complete and arbitrage-free.
As the riskless rate derived  for this perpetual derivative is applicable to the chosen market class,
and as it mimics the efforts of shadow bank portfolio managers trying to create riskless rates using their available assets,
they refer to the computed riskless rate as the shadow riskless rate (SRR).

To date, there has been no empirical investigation associated with this theoretical development.
The aim of the present paper is to develop a practical numerical implementation of the approach proposed in
\cite{Rachev_2017} under the correlated GBM assumption
(the simplest model under which to develop, test and improve the needed numerical methods).
For $N$ assets whose prices are driven by $N-1$ Brownian motions,
the theoretical method for deriving an SRR, as introduced by \cite{Rachev_2017}, is summarized in
Section \ref{sec:model}.
Theoretically, the SRR is obtained as the unique solution of a linear system.
In Section~\ref{sec:calibration}, given a window of empirical historical returns for the assets,
we describe how to use principal component analysis to model the actions of the Brownian motions.
Based on the PCA analysis, we derive two methods for generating the components of the coefficient
matrix required for that linear system.
In Section~\ref{sec:timeseries} we develop the moving window method to generate a time series
of SRR values.
We demonstrate numerically unstable behavior in the time series arising when the coefficient matrix rapidly changes
condition number.
In Sections~\ref{sec:reg} and \ref{sec:re_reg} we develop an approach to provide
a regularized solution when the condition number of the linear system is detected to be changing significantly.
Section~\ref{sec:compare} is devoted to comparing solutions obtained using the two different methods for
generating the coefficient matrix.
As the SRR is based upon the drift component of a state-price deflator,
an additional outcome of this analysis is the estimation of the total volatility of this deflator.
The numerical estimation of this is discussed in Section~\ref{sec:vol_pi}.

Application of the numerical method to empirical datasets is described in Section~\ref{sec:empirical}.
Section \ref{sec:type} computes estimates  of the SRR for four different groups of $N$ stocks as well as
a group of $N$ exchange-traded funds.
Section \ref{sec:size} addresses the effect of the number $N$ of stocks in the group.
Section \ref{sec:spd} turns attention to the behavior of the state-price deflator, plotting the trajectory
over time of its drift and volatility.
Section \ref{sec:disc} concludes the paper with a discussion of future research directions and a proposal
for an alternate empirical definition of a riskless rate and its proxy asset.

\section{Method}\label{sec:method}

\subsection{Derivation of the Shadow Riskless Rate} \label{sec:model}

Consider a portfolio of $N$ risky assets  ${\cal S}^{(j)}$, $j = 1,\dots,N$, $N \ge 2$.
Asset ${\cal S}^{(j)}$ has price $S_{jt}$, $t \ge 0$ whose dynamics is driven by
$N-1$ Brownian motions $W_{kt}$, $k = 1, ..., N-1$,
and is given by the stochastic differential equation
\begin{equation}\label{eq:GBM}
	dS_{jt} = \mu_{j} S_{jt} dt + S_{jt} \sum_{k=1}^{N-1} \sigma_{jk}dW_{kt},
	\quad t \ge 0,\ \ \mu_j > 0, \ \ \sigma_j > 0, \quad j=1,  \dots,N,
\end{equation} 
with  the initial conditions $S_{jt_{0}}=S_{j0} > 0$, $j = 1, ..., N$. 
We define the mean return vector  $\mu = [ \mu_1, \dots, \mu_N ]$ and the
variance--covariance matrix $\Sigma \Sigma^T$, where $\Sigma_{jk}=\sigma_{jk}$,  $ j=1,\dots,N$,  $k=1,\dots,N-1.$
The Brownian motions generate a filtered probability space
$(\Omega,{\cal F},\mathbb{F}=\{{\cal F}_t, t \ge0\},\mathbb{P})$ representing the natural world.
The market of risky assets ${\cal S}^{(j)}$, $j = 1,\dots,N$, having price dynamics described by (\ref{eq:GBM})
will be complete if and only if there exists a unique  state-price deflator
$\pi_t$, $t \ge 0$, on $\mathbb{P}$ \cite[section 6D]{Duffie_2001} with dynamics given by the It\^{o} process
\begin{equation}\label{eq:Pi}
	d\pi_t = \mu_\pi \pi_t dt + \pi_t \sum_{k=1}^{N-1} \sigma_{\pi k}dW_{kt}, \quad t \ge 0.
\end{equation} 

The existence and uniqueness of $\pi_t$ is the statement that each deflated process
$S_{jt} \pi_t$, $j = 1, ..., N$, be a $\mathbb{P}-$martingale.
This leads to the requirement that the linear system
\begin{equation}\label{eq:Pi2}
  	\mu_j + \mu_\pi + \sum_{k=1}^{N-1} \sigma_{jk} \sigma_{\pi k}  =  0, \quad  j=1,  \dots,N,
\end{equation} 
have a unique solution for $\mu_\pi$ and $\sigma_{\pi k}$, $k = 1, ..., N-1$.
In matrix notation, \eqref{eq:Pi2} can be written
\begin{equation}\label{eq:linsys}
	\Phi x = \mu,
\end{equation}
where $\mu$ is the column vector $[\mu_1, ..., \mu_N]^{\text{T}}$,
$x$ is the solution vector $[-\mu_\pi, \sigma_{\pi 1}, ..., \sigma_{\pi (N-1)}]^{\text{T}}$,
and the matrix $\Phi = [ \one_N, - \Sigma]$ with $\one_{N}$ being the $( N \times 1)$ column vector composed of ones.
The SRR $\nu$ is given by $\nu = -\mu_{\pi}$ \citep{Rachev_2017}.
From \eqref{eq:linsys}, it has the analytic solution
\begin{equation}\label{eq:SRR}
	\nu = \frac{\det \Phi_\mu }{ \det \Phi},
\end{equation} 
where $\Phi_\mu = [ \mu, - \Sigma]$.
\cite{Rachev_2017} construct the price process $B_t$ of a tradeable, perpetual European contingency contract
${\cal B}$ that serves the role of a riskless bond in the arbitrage-free, complete market $({\cal S}_1, ..., {\cal S}_N, {\cal B})$.
The price dynamics of ${\cal B}$ obeys
\begin{equation}\label{eq:dB}
	dB_t = \nu B_t dt.
\end{equation}

The analytic solution for each standard deviation $\sigma_{\pi k}$, $k = 1, ..., N-1$,
can be expressed similarly as a ratio of determinants.
Thus while $\mu_\pi = -\nu$ measures the drift component of the state-price deflator,
the vector $[ \sigma_{\pi 1} , \dots, \sigma_{\pi (N-1)} ]$ estimates the total variance $ \sigma^2_\pi$
of the state-price deflator,
\begin{equation}\label{eq:sigma_pi}
	\sigma^2_\pi = \sum_{k=1}^{N-1} (\sigma_{\pi k})^2 .
\end{equation}
We shall refer to $\sigma_\pi = \sqrt{ \sigma_\pi^2 }$ as the (total) volatility of the state-price deflator.

We make some critical observations of the SRR and the total volatility of the state-price deflator
by briefly describing the solution when $N = 2$.
(See, also, \cite{Rachev_2017}.)
Equations \eqref{eq:GBM} become
\begin{equation}\label{eq:GBM_12}
	\begin{aligned}
		dS_{1t} &= \mu_1 S_{1t} dt + \sigma_1 S_{1t} dW_t, \quad \mu_1 > 0, \ \ \sigma_1 > 0, 	\quad t \ge 0,\\
		dS_{2t} &= \mu_2 S_{2t} dt + \sigma_2 S_{2t} dW_t, \quad \mu_2 > 0, \ \ \sigma_2 > 0, 	\quad t \ge 0;
	\end{aligned}
\end{equation}
equation \eqref{eq:Pi} becomes
\begin{equation}\label{eq:Pi_12}
	d\pi_t = \mu_\pi \pi_t dt + \sigma_\pi \pi_t dW_t, \quad t \ge 0;
\end{equation}
and the linear system \eqref{eq:Pi2} can be written as
\begin{equation}\label{eq:Pi2_12}
  	\ -\mu_\pi  =  \mu_1 + \sigma_1 \sigma_\pi  =  \mu_2 + \sigma_2 \sigma_\pi,
\end{equation} 
leading to the solutions
\begin{subequations}
\begin{align}
	\nu = -\mu_\pi &= \frac{ \mu_1 \sigma_2 - \mu_2 \sigma_1 }{ \sigma_2 - \sigma_1  }, \label{eq:soln_12a}\\
	\sigma_\pi &= \frac{ \mu_1 - \mu_2 }{ \sigma_2 - \sigma_1 }.\label{eq:soln_12b}
\end{align}
\end{subequations}
Equation \eqref{eq:Pi2_12} provides the useful identity between the market prices of risk for the two assets,
\begin{equation}\label{eq:mpor}
	\frac{ \mu_1 - \nu }{ \sigma_1 }  =  - \sigma_\pi = \frac{ \mu_2 - \nu }{ \sigma_2 } ,
\end{equation}
reinforcing the view of $\nu$ as a riskless rate.
Without loss of generality, we can assume $\Delta \sigma = \sigma_2 - \sigma_1 > 0$.
Defining $\Delta \mu = \mu_2 - \mu_1$, \eqref{eq:soln_12b} and \eqref{eq:soln_12a} can be written
\begin{equation}\label{eq:sign}
	\sigma_\pi = -\frac{ \Delta \mu }{ \Delta \sigma }, \qquad
	\nu = -\mu_\pi =  \mu_1 \left( 1 + \frac{ \sigma_\pi }{ \mu_1 / \sigma_1  } \right).
\end{equation}
As $\mu_1 > 0$, $\sigma_1 > 0$ and $\Delta \sigma > 0$, we see that $\sigma_\pi$ can either be positive or
negative depending on the sign of $\Delta \mu$.
If $\sigma_\pi$ is positive, then $\nu$ is positive (and $\mu_\pi$ is negative).
If $\sigma_\pi$ is negative, $\nu$ can be either positive or negative
depending on the magnitude of the ratio $ \sigma_\pi / (\mu_1 / \sigma_1)$.
For $N > 2$ risky assets, the determination of the sign of the solution \eqref{eq:SRR} is more complicated,
but can be either positive or negative.

\subsection{Calibration to Historical Data} \label{sec:calibration}

In practice, estimation of the SRR via \eqref{eq:linsys} involves the estimation of the components $\mu_j$ and $\sigma_{jk}$ 
of $\mu$ and $\Sigma$ in the discrete setting $dt \approx \Delta t $.
In discrete time, the log return $r_j$ for asset $j$ will have instantaneous mean $\mu_j$
and instantaneous standard deviation vector $[\sigma_{j1}, \dots, \sigma_{j (N-1)}]$.
Estimation of these components requires a method to retrieve the $N-1$ Brownian motions that drive the
uncertainty in the return dynamics.
As the distributional assumption embodied in \eqref{eq:GBM} is multivariate geometric Gaussian,
we use principal component analysis (PCA) to do this.
A brief summary of PCA is provided in the appendix.
Here we employ the matrix row and column vector notation introduced in the appendix.

Let $R$ denote the $M \times N $ matrix of daily log-returns $r_{mj}$ realized by the $N$ risky assets over $M$ historical trading days 
(i.e., $\Delta t=1$).
We estimate the mean log-return value $\mu_j$ of each asset $j$ from the historical data.
The historical log-returns can then be adjusted by subtracting historical mean values,
producing a modified matrix $R$ which has zero mean in each column.

Applying PCA to $R$ results in $N$ ordered eigenvector-eigenvalue pairs.
As the multivariate price process is driven by multivariate geometric Brownian motion,
the returns $r_{mj}$ are distributed in a hyper-ellipsoid in $\mathbb{R}^N$.
The column eigenvectors $w_{\circ 1},\dots,w_{\circ N}$ from the PCA,
arranged in order of the variance they explain,
represent the principal axes of this hyper-ellipsoid.
The components $p_{mj}$, $m = 1, ..., M$, of the principal component column vector $p_{\circ j} = R w_{\circ j}$
describe the marginal distribution of the vector of correlated returns $(r_{m1}, ..., r_{mN})$
on the $j$'th principal axis of the hyper-ellipsoid.
As there are $N-1$ Brownian motions driving the return distribution,
we make the approximation that the specific actions of these $N-1$ Brownian motions on the observed dataset
are described by the principal components observed along the first $N-1$ principal axes, $w_{\circ j}$, $j = 1, ..., N-1$.
(We ignore the distribution along $w_{\circ N}$, as this principal axes explains the smallest variance of the observed dataset.)
Based on \eqref{eq:pca_cov}, we have the approximation
\begin{equation}\label{eq:sig_jk}
	\sigma_{jk} = \sqrt{\lambda_k} w_{jk}, \quad j = 1, ..., N, \quad k = 1, ..., N-1.
\end{equation}

There is a second method to approximate the values $\sigma_{jk}$.
Again, assuming that the actions of the $N-1$ Brownian motions on the observed dataset are described by
the distributions of the principal components observed along the first $N-1$ principal axes, we can approximate
\begin{equation}\label{eq:reg_sig_jk}
	r_{mj} - \text{E}_m[r_{mj}] = \sum_{k=1}^{N-1}  \sigma_{jk} \bar{P}_{ik}, \quad m = 1, ..., M, \quad j = 1, ..., N,
\end{equation}
where
\begin{equation}\label{eq:Pbar}
	\bar{P}_{ik} = \frac{ P_{ik} - \text{E}_i [P_{ik}] }{ \sqrt{ \text{Var}_i [P_{ik}] } },
\end{equation}
with $\text{E}_m[r_{mj}]$ and $\text{E}_i [P_{ik}]$ denoting the average of the indicated column,
and $\text{Var}_i [P_{ik}]$ denoting the column variance.
Solution of the values $\sigma_{jk}$ in \eqref{eq:reg_sig_jk} is obtained via linear regression.\footnote{
	Attempts to use robust linear regression with M-estimation produced a $\Sigma$ matrix whose
	condition number, and freqency of change of condition number, was even worse.
}

\subsection{Computation of a Time Series for the SRR} \label{sec:timeseries}

The calibration described in Section \ref{sec:calibration} requires a historical window (of size $M$ days)
to compute a single value of the SRR $\nu$ via \eqref{eq:linsys}.
Using a standard moving window procedure, we can develop a time series  of historical SSR values.
Given a group of $N$ assets, we compute a value for $\nu_t$ for date $t$ as follows.
\begin{enumerate}
\item Assemble the log-return matrix $R_t$ for these $N$ assets over the historical window $\{ t-(M-1), \dots,  t-1, t \}$
	of $M$ trading days.
\item  Estimate the vector $\mu_t$ from the historical data and subtract the respective mean values
from each column of the matrix $R_t$.
\item Perform a PCA of $R_t$, producing the ordered eigenvalue and eigenvector pairs, $\lambda_j^{(t)}$, $w_{\circ j}^{(t)}$, $j = 1, ..., N$.
\item Generate the matrix $\Sigma_t$ composed of column vectors $\sigma_{\circ k}^{(t)}$, $k = 1, ..., N-1$
using either (4.i) the direct solution \eqref{eq:sig_jk} or (4.ii) the regression solution to \eqref{eq:reg_sig_jk}.
\item Form $\Phi_t$. 
\item While $\nu_t$ can be obtained analytically from (\ref{eq:SRR})
and the standard deviations $\sigma_{\pi k}$ from analogous determinant ratios,
numerically it is more efficient and stable to solve \eqref{eq:linsys} using LU decomposition with pivoting.
\end{enumerate}
Steps 1 through 6 generate the time series
\begin{equation}\label{eq:srr+ts}
	\nu_t = x_{1,t}, \qquad \sigma_{\pi k, t} = x_{k,t}, \qquad k = 2, ..., N.
\end{equation}

Step 6 however is vulnerable to numerical instability,
specifically the sensitivity of the linear system (\ref{eq:linsys}) to the condition number of the matrix $\Phi_t$.
The condition number is
\begin{equation} \label{eq:cond_number}
	\kappa_t  = \frac{ \lambda_H (\Phi_t) } { \lambda_L (\Phi_t) },
\end{equation}
where $\lambda_H (\Phi_t)$ and $\lambda_L (\Phi_t)$ denote, respectively, 
the highest and lowest absolute values of the eigenvalues of $\Phi_t$.
A large condition number indicates that the matrix is close to being singular with $\text{det}\ \Phi_t \approx 0$ \citep{Cheney_2012}. 

\begin{figure}[htb]
	\centering
	\includegraphics[width=\textwidth]{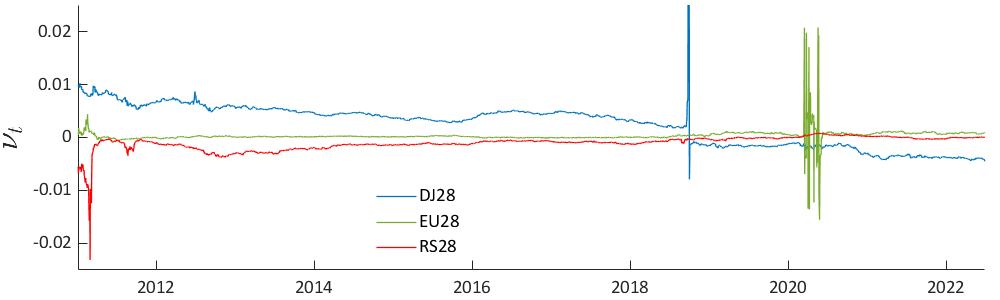} 
	\caption{Plots of the SRR $\nu_t$ computed using steps 1 through 6, with \eqref{eq:sig_jk} used in step 4.
		Note that the large positive spike in the plotted time series for DJ28 has been truncated to 11\% of its true height.
	}
	\label{fig:det_soln}	
\end{figure}
\begin{figure}[htb]
	\centering
	\includegraphics[width=0.5\textwidth]{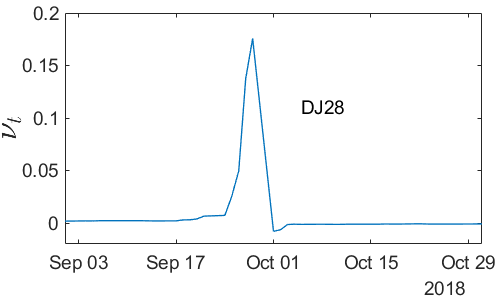} 
	\caption{Plots of the SRR $\nu_t$ computed using \eqref{eq:SRR}.}
	\label{fig:nu_dj_0918}	
\end{figure}

Fig.~\ref{fig:det_soln} shows three example time series $\nu_t$ computed using the above procedure,
with the elements of vector $\sigma_{\pi t}$ computed using  \eqref{eq:sig_jk}.
The three time  series use log-returns from $N= 28$ stocks selected from components of the Russell 3000 (RS28),
the STOXX Europe 600 (EU28), and the Dow Jones (DJ28) indices.
(The selection of these stocks is discussed in Section~\ref{sec:type}.)
A historical window of $M = 2500$ days was used.
The time series illustrate the following issues.
\begin{enumerate}
\item[Ia.] The presence of ``spikes'' indicating periods when the matrix $\Phi_t$ changes condition number rapidly with time.
\item[Ib.] The fact that the investigation of the time series may begin during a period of rapidly changing
condition number (as in the RS28 time series).
\item[Ic.] During ``normal'' periods (when the daily change in condition number $\Phi_t$ is random and ``small'')
the value of $\nu_t$ has random behavior.
\item[Id.] There may be long term trends in $\nu_t$ (seen in all three, but most notably DJ28).
\item[Ie.] There may be discontinuous changes in the ``baseline'' value of $\nu_t$ associated with rapid
changes in condition number (as in the DJ28 time series).
\end{enumerate}
As illustrated in Fig.~\ref{fig:nu_dj_0918}, these spikes always occur over a multiday period
-- the prominent spike in the DJ28 time series occurs over the period 25 Sept. to 1 Oct., 2018.
Thus regularization techniques that attempt to set $\nu_t = \nu_{t-1}$
when the condition number of $\Phi_t$ is flagged as  changing significantly
will generally not work.

\subsection{Regularization of the Matrix $\Phi_t$} \label{sec:reg}

Developing a reliable estimate for $\nu_t$ requires
(RE.i) a method for identifying when the condition number of the matrix $\Phi_t$ changes significantly
and
(RE.ii) implementing a procedure to reduce the change appropriately at those times.
The estimation procedure must preserve behaviors Ic and Id while appropriately addressing Ia, Ib and Ie.
Identifying when the condition number $\kappa_t$ of $\Phi_t$ changes significantly requires a characterization of
``what value of $\kappa_t - \kappa_{t-1}$ is too large?''

There are a variety of smoothing methods possible for addressing point RE.ii that indirectly control
the rate of change of the condition number.
One approach is to solve \eqref{eq:linsys} as a least squares minimization problem with penalty (regularization) terms,
\begin{equation}\label{eq:regLS}
	\bar{x}_t = \argmin_z \left( ||\Phi_t z - \mu_t||^2_2 +  || \Gamma_t z||^2_2 \right),
\end{equation}
for some choice of the matrix $\Gamma_t$, which defines the impact of  the  regularization  on the solution.\footnote{
	Solution of \eqref{eq:regLS} is equivalent to finding the solution $x_t$ that minimizes the $L_2$-norm
	of the error vector $\epsilon_t$ of the regression problem $\mu_t =  \Phi_t x_t + \epsilon_t$.
	Approach \eqref{eq:regLS} is know as Tikhonov regularization \citep{Tikhonov_1965} when applied to integral
	equations,
	and ridge regression \citep{Hoerl_1970} when applied to finite-dimensional regression problems.
}
An example of such a regularization is the minimization
\begin{equation}\label{eq:OptLS}
	\bar{x}_t = \argmin_{z_1, ..., z_N} \left[ ||\Phi_t z - \mu_t||^2_2 + \gamma_{1t} ( z_1+\mu_{\pi (t-1)} )^2
		 + \gamma_{2t} \left( \sum_{k=2}^N z_k^2 \right) \right ],
\end{equation}
where $\mu_{\pi (t-1)} $ is the solution of the SRR computed for the previous day
and the coefficients $\gamma_{1t}, \gamma_{2t}$ are used to weight the regularization components.
The first penalty term controls the rate of change of the SRR;
the second penalty term penalizes solutions with large variance (\ref{eq:sigma_pi}).
The penalty terms must be time-dependent to avoid regularization when $\kappa_t$ is not changing significantly.
We have investigated approach \eqref{eq:OptLS}.
Modeling $\gamma_{1t}$ and $\gamma_{2t}$ led us to a four-parameter model whose parameters required
adjustment with choice of asset universe.

We propose instead a method which has several advantages.
It addresses points RE.i and RE.ii simultaneously;
it provides an understanding of what is controlling $\kappa_t$;
it involves a single regularization parameter;
and it is computationally efficient.
Consider the singular value decomposition of the matrix $\Phi_t$,
\begin{equation}\label{eq:SVD}
	\Phi_t = U_t D_t V_t^{\text{T}}
\end{equation}
where $U_t$ and $V_t$ are orthogonal matrices in $R^{N \times N}$ and $D_t = \text{diag}(d_1^{(t)}, ..., d_n^{(t)})$
consists of the singular values ordered such that $d_1^{(t)} > d_2^{(t)} ... > d_n^{(t)}$.
Solution of \eqref{eq:linsys} is equivalent to solving
\begin{subequations}
\begin{align}
	y &= U_t^{\text{T}} \mu, \label{eq:SVDa}\\
	z &= D_t^{-1} y,  \label{eq:SVDb}\\
	x_t &= V_t z.  \label{eq:SVDc}
\end{align}
\end{subequations}
As the condition number of an orthogonal matrix is unity,
the ill-conditioning in the solution $x_t$ arises from \eqref{eq:SVDb},
specifically, as we show next, from $z_n = y_n / d_n^{(t)}$.

\begin{figure}[htb]
	\centering
	\includegraphics[width=\textwidth]{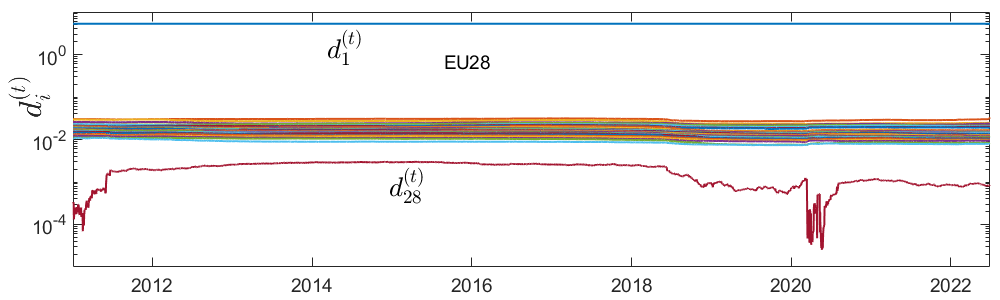}
	\includegraphics[width=\textwidth]{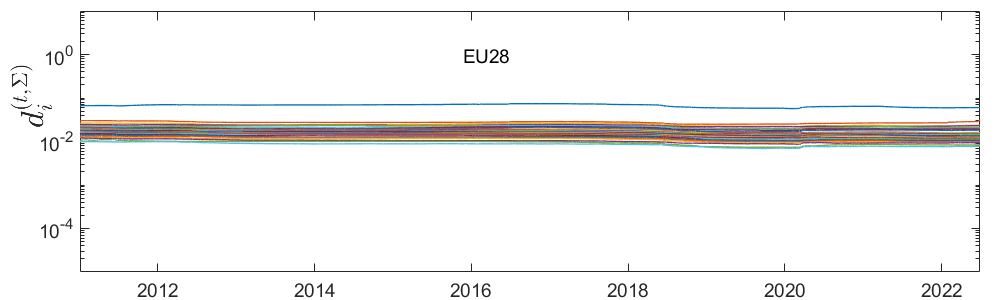}
	\caption{Time series of (top) the singular values $d_j^{(t)}$, $j = 1, ..., 28$ of the matrix $\Phi_t$ 
		and (bottom) the singular values $d_j^{(t,\Sigma)}$, $j = 1, ..., 27$ of the matrix $\Sigma_t$
		 for the RS28 dataset.}
	\label{fig:nu_EU}	
\end{figure}
Fig.~\ref{fig:nu_EU} plots the time series of singular values $d_j^{(t)}$, $j = 1, ..., 28$, corresponding to the calculation of
$\nu_t$ values for EU28 shown in  Fig.~\ref{fig:det_soln}.
Note the large separations between  the group of singular values  $d_j^{(t)}$, $j = 2, ..., 27$,
and the largest $d_1^{(t)}$ and smallest $d_{28}^{(t)}$ singular values.
(This is a consistent finding in all the cases studied in Section~\ref{sec:empirical}.)
The separation of $d_1^{(t)}$ from the remaining singular values reflects the size difference between the first column
of values $\one_N$ of the matrix $\Phi_t$ and the remaining columns of the submatrix $-\Sigma_t$ of $\Phi_t$.
Also plotted in Fig.~\ref{fig:nu_EU} is the time series of (non-zero) singular values $d_j^{(t,\Sigma)}$, $j = 1, ..., 27$
of the submatrix $\Sigma_t$.
The submatrix $\Sigma_t$ is responsible for the intermediate group of singular values $d_j^{(t)}$, $j = 2, ..., 27$, seen in $\Phi_t$.

\begin{figure}[htb]
	\centering
	\includegraphics[width=\textwidth]{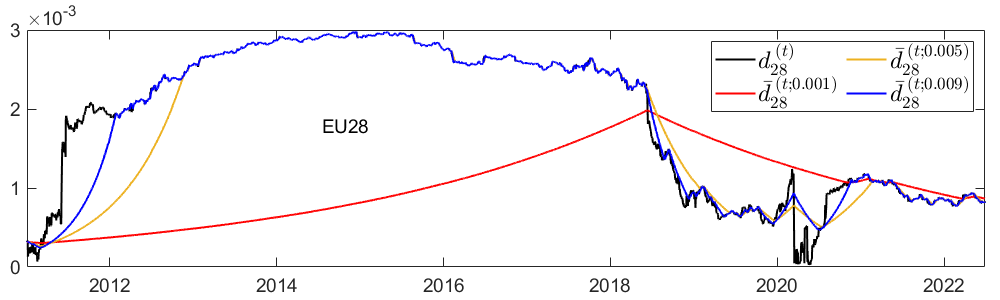}
	\includegraphics[width=\textwidth]{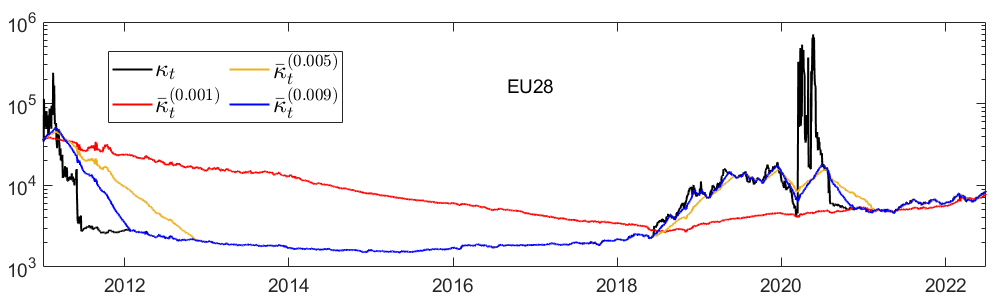}
	\caption{The EU28 time series (top) $d_{28}^{(t)}$ and $\bar{d}_{28}^{(t;\epsilon)}$;
	and (bottom) $\kappa_t$ and $\bar{\kappa}_t^{(\epsilon)}$
	for the regularization values $\epsilon=\{0.001, 0.005, 0.009\}$. }
	\label{fig:dnu_bar_EU}	
\end{figure}

Variation in the smallest singular value $d_{28}^{(t)}$ controls the condition number of $\Phi_t$ and the spiking behavior
shown in $\nu_t$.
It results when the first column of $\Phi_t$, the vector $\one_N$, becomes close to the space spanned by the columns of
$\Sigma_t$.
We therefore developed an algorithm to control the time rate of change of the smallest singular value $d_n^{(t)}$ of $\Phi_t$.
As the singular values are always positive, we regularize the smallest singular value as
\begin{equation}\label{eq:eps_reg}
	\begin{aligned}
	\bar{d}_n^{\;(t,\epsilon)} & = 
	\begin{cases}
		\min \left( d_n^{(t)}, (1+\epsilon) \bar{d}_n^{\;(t-1,\epsilon)} \right) 
				& \textrm{ if } d_n^{(t)} \ge \bar{d}_n^{\;(t-1,\epsilon)}, \\
		\max \left( d_n^{(t)}, (1-\epsilon) \bar{d}_n^{\;(t-1,\epsilon)}  \right)
				& \textrm{ if } d_n^{(t)} < \bar{d}_n^{\;(t-1,\epsilon)}, 
	\end{cases} \\
	\bar{d}_n^{\;(0,\epsilon)} &= d_n^{(0)}.
	\end{aligned}
\end{equation}
Fig.~\ref{fig:dnu_bar_EU} illustrates the sensitivity of the regularization of $d_n^{(t)}$ to the value $\epsilon$;
too small a value of $\epsilon$ leads to too much regularization,
too large a value does not adequately smooth the spikes in $d_n^{(t)}$.

Let  $\overline{D}_t^{(\epsilon)}$ denote the matrix $D$ in \eqref{eq:SVD} having the singular value $d_n^{(t)}$
replaced by the regularized value $\bar{d}_n^{\;(t;\epsilon)}$.
Let $\bar{\kappa}_t^{(\epsilon)}$ denote the condition number of the regularized matrix
$\overline{\Phi}_t^{(\epsilon)} \equiv U_t \overline{D}_t^{(\epsilon)} V_t^{\textrm{T}}$.
Fig.~\ref{fig:dnu_bar_EU} compares $\kappa_t$ against $\bar{\kappa}_t^{(\epsilon)}$ for
$\epsilon \in \{0.001, 0.005, 0.009\}$.
Of course, the behavior of $\bar{\kappa}_t^{(\epsilon)}$ relative to $\kappa_t$ must reflect
the behavior of $\bar{d}_n^{\;(t,\epsilon)}$ relative to $d_n^{\;(t)}$.

\begin{figure}[htb]
	\centering
	\includegraphics[width=\textwidth]{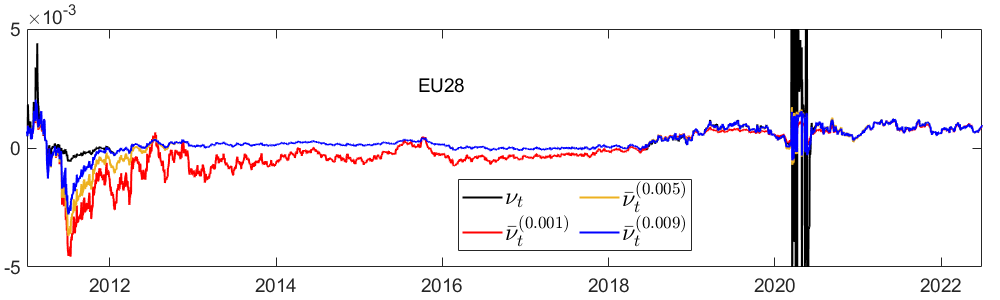}
	\includegraphics[width=\textwidth]{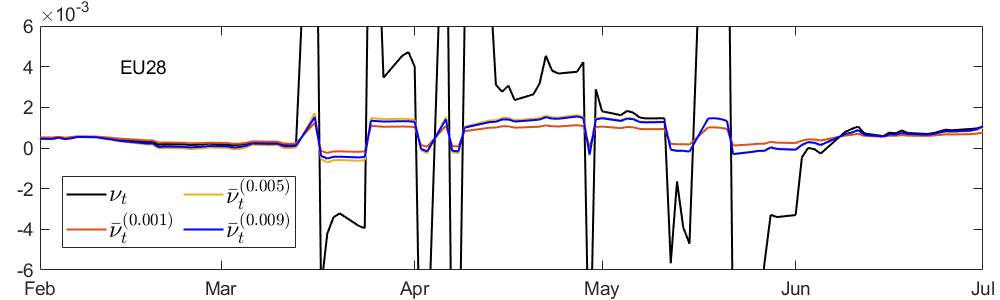}
	\caption{(top) the EU28 times series $\nu_t$ and $\bar{\nu}_t^{(\epsilon)}$ and
	(bottom) a detailed view, over the period 01 Feb., 2020 through 01 July, 2020, of these two series
	for the regularization values $\epsilon=\{0.001, 0.005, 0.009\}$. }
	\label{fig:nu_bar_EU_2020}	
\end{figure}

Let $\bar{\nu}_t^{(\epsilon)}$ denote the solution obtained
from \eqref{eq:SVDa}--\eqref{eq:SVDc} using $\overline{D}_t^{(\epsilon)}$ in \eqref{eq:SVDb}.
Fig.~\ref{fig:nu_bar_EU_2020} compares the results of  $\nu_t$ against $\bar{\nu}_t^{(\epsilon)}$
for $\epsilon \in \{0.001, 0.005, 0.009\}$.
The results for the EU28 selected stocks indicate that, of the three, the choice $\epsilon = 0.009$ provides the best regularization.
The plot also indicates the ease with which a choice for $\epsilon$ can be made for a particular selection of stocks.
The figure also shows detail of the behavior of the regularized values $\bar{\nu}_t^{(\epsilon)}$
for the period 01 Feb., 2020 through 01 July, 2020.
The regularized values ``respond'' to the extreme volatility expressed by $\nu_t$, but in a highly muted manner.
There is no visible difference between the $\epsilon = 0.005$ and  $\epsilon = 0.009$ regularizations over this
time period.

The EU28 dataset is specifically chosen to illustrate two features reflecting the fact that the condition number of
$\Phi_t$ can change its ``baseline'' value over time.
\begin{enumerate}
\item As the EU28 time series is first sampled at the start of 2011 during a period over which $\kappa_t$ is rapidly changing,
and the baseline of $\nu_t$ appears to be changing,
the strength of the regularization determines how soon the regularized values can ``settle''  to the appropriate condition number and new baseline.
In contrast, if the time series is first sampled during a period in which $\kappa_t$ is not changing significantly,
this initial ``transient'' behavior seen in the EU28 dataset would not appear.
\item Beginning in 2018, $\kappa_t$ again begins to change, with $d_{28}^{(t)}$ decreasing from a
baseline value of roughly 0.0025 to  a new baseline of roughly 0.001 at the start of 2021.
The regularization must accommodate such a baseline change without ``looking into the future'',
but strongly correct for extreme spiking behavior such as seen in 2020.
\end{enumerate}

Although the value $\epsilon = 0.009$ produces the best result in Fig.~\ref{fig:dnu_bar_EU}, 
it produces only a slight improvement over the choice  $\epsilon = 0.005$.
For the other asset groups discussed in Section~\ref{sec:empirical}, $\epsilon = 0.005$ produces the overall
best results.
For consistency, we therefore proceed with the value  $\epsilon = 0.005$.

\subsection{Secondary Regularization of the SRR} \label{sec:re_reg}

\begin{figure}[htb]
	\centering
	\includegraphics[width=\textwidth]{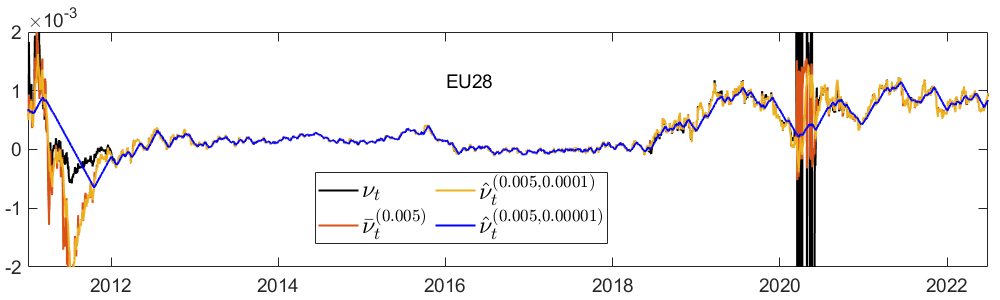}
	\caption{The EU28 time series $\nu_t$, $\bar{\nu}_t^{(0.005)}$ and $\hat{\nu}_t^{(0.005,\delta_\nu)}$
	for the regularization values $\delta_\nu \in \{10^{-4}, 10^{-5}\}$. }
	\label{fig:dnu_bar_hat_EU}	
\end{figure}

The regularization \eqref{eq:eps_reg} can be repeated directly on the time series $\bar{\nu}_t^{(\epsilon)}$, 
producing a further smoothed SRR value
\begin{equation}\label{eq:del_reg_nu}
	\begin{aligned}
	\hat{\nu}_t^{(\epsilon,\delta_\nu)} & = 
	\begin{cases}
		\min \left( \bar{\nu}_t^{(\epsilon)}, (1+\delta_\nu) \hat{\nu}_{t-1}^{(\epsilon,\delta_\nu)} \right)
			& \textrm{ if } \bar{\nu}_t^{(\epsilon)} \ge \hat{\nu}_{t-1}^{(\epsilon,\delta_\nu)}, \\
		\max \left(  \bar{\nu}_t^{(\epsilon)}, (1-\delta_\nu)  \hat{\nu}_{t-1}^{(\epsilon,\delta_\nu)}  \right)
			& \textrm{ if } \bar{\nu}_t^{(\epsilon)} < \hat{\nu}_{t-1}^{(\epsilon,\delta_\nu)}, 
	\end{cases} \\
	\hat{\nu}_0^{(\epsilon,\delta_\nu)} &= \bar{\nu}_0^{(\epsilon)}.
	\end{aligned}
\end{equation}
Fig.~\ref{fig:dnu_bar_hat_EU} shows the effect of this secondary smoothing using the values
$\delta_\nu \in \{10^{-4}, 10^{-5}\}$.
As $\delta_\nu$ decreases further, the time series $\hat{\nu}_t^{(\epsilon,\delta_\nu)}$ becomes increasingly linear,
losing time dependent details.
We therefore proceed using the secondary smoothing parameter value $\delta_\nu = 10^{-5}$.

\subsection{Comparison of Solution Methods \eqref{eq:sig_jk} and \eqref{eq:reg_sig_jk}} \label{sec:compare}
Equations \eqref{eq:sig_jk} and \eqref{eq:reg_sig_jk} present alternate methods for approximating the entries
$\sigma_jk$ of the matrix $\Sigma$.
The results discussed above have used \eqref{eq:sig_jk}.
Here we briefly illustrate the differences observed in using the two methods using the EU28 dataset.

\begin{figure}[htb]
	\centering
	\includegraphics[width=\textwidth]{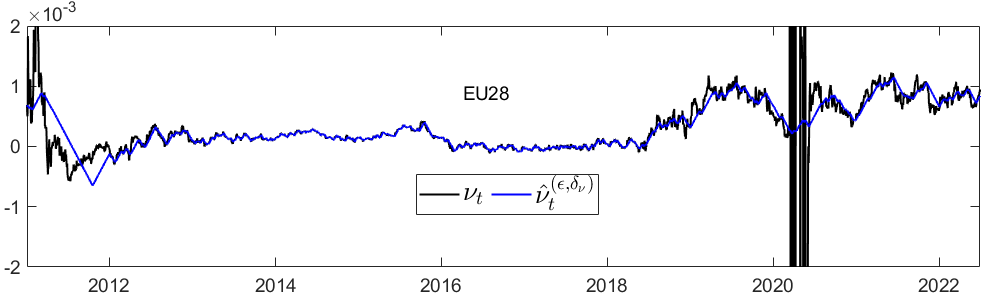}
	\includegraphics[width=\textwidth]{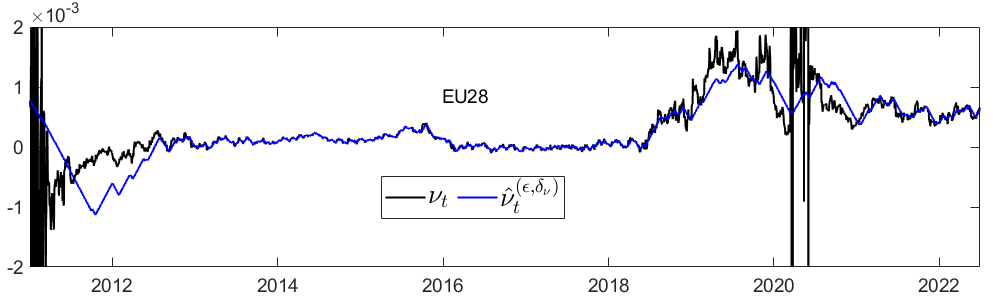}
	\caption{The EU28 time series $\nu_t$ and $\hat{\nu}_t^{(\epsilon,\delta_\nu)}$
		for $\sigma_{jk}$ obtained via (top) \eqref{eq:sig_jk} and (bottom) \eqref{eq:reg_sig_jk}.}
	\label{fig:v9v11_nu_EU}	
\end{figure}
Fig.~\ref{fig:v9v11_nu_EU} compares the EU28 time series $\nu_t$ and $\hat{\nu}_t^{(\epsilon,\delta_\nu)}$
for $\sigma_{jk}$ obtained via \eqref{eq:sig_jk} and \eqref{eq:reg_sig_jk}.
Use of  \eqref{eq:reg_sig_jk} produces an unregularized matrix $\Phi_t$ that is generally has significant changes in
$\kappa_t$ more frequently than that obtained using  \eqref{eq:sig_jk}.
As a result, we applied the regularization \eqref{eq:eps_reg} to each singular value $d_k^{(t)}$, $k = 1, ..., N$,
producing the regularized matrix $\overline{D}_t^{(\epsilon)}$ and corresponding regularized solution
$\bar{\nu}_t^{(\epsilon)}$.
Secondary regularization via \eqref{eq:del_reg_nu} produces the solution $\hat{\nu}_t^{(\epsilon,\delta_\nu)}$.
The secondary regularized solutions are comparable between the two methods.
We have found that, depending on the dataset, either of the two methods may produce secondarily regularized solutions
having smaller variation over time.

\subsection{Estimation and Regularization of the Total Volatility $\sigma_{\pi,t}$} \label{sec:vol_pi}

As noted in Section~\ref{sec:model}, solution of \eqref{eq:linsys} enables a determination of the total
volatility $\sigma_{\pi,t}$ of the state-price deflator via \eqref{eq:sigma_pi}.
Let $ \bar{\sigma}_{\pi k,t}^{(\epsilon)}$, $k = 1, ..., N-1$, denote the solutions obtained
from \eqref{eq:SVDa}--\eqref{eq:SVDc} using $\overline{D}_t^{(\epsilon)}$ in \eqref{eq:SVDb}.
This produces the regularized total volatility
$\bar{\sigma}_\pi^{(\epsilon)} = \sqrt{\sum_{k=1}^{N-1} (\bar{\sigma}_{\pi k,t}^{(\epsilon)})^2}$.
A secondary regularization can also be applied directly to the time series $\bar{\sigma}_{\pi k, t}^{(\epsilon)}$,
$k = 1, ..., N-1$, producing
\begin{equation}\label{eq:del_reg_sigma}
	\begin{aligned}
	\hat{\sigma}_{\pi k,t}^{(\epsilon,\delta_\sigma)} & = 
	\begin{cases}
		\min \left( \bar{\sigma}_{\pi k,t}^{(\epsilon)}, (1+\delta_\sigma) \hat{\sigma}_{\pi k, t-1}^{(\epsilon,\delta_\sigma)} \right)
			& \textrm{ if }	\bar{\sigma}_{\pi k,t}^{(\epsilon)} \ge \hat{\sigma}_{\pi k,t-1}^{(\epsilon,\delta_\sigma)}, \\
		\max \left(  \bar{\sigma}_{\pi k,t}^{(\epsilon)}, (1-\delta_\sigma)  \hat{\sigma}_{\pi k, t-1}^{(\epsilon,\delta_\sigma)}  \right)
			& \textrm{ if }	\bar{\sigma}_{\pi k,t}^{(\epsilon)} < \hat{\sigma}_{\pi k,t-1}^{(\epsilon,\delta_\sigma)}, 
	\end{cases} \\
	\hat{\sigma}_{\pi k,0}^{(\epsilon,\delta_\sigma)} &= \bar{\sigma}_{\pi k,0}^{(\epsilon)}.
	\end{aligned}
\end{equation}
The difference in the scale of the values $\nu_t$ and $\sigma_{\pi k,t}$ necessitates that
$\delta_\sigma > \delta_\nu$.
We found the value $\delta_\sigma = 10^{-3}$ provided adequate smoothing.

\begin{figure}[htb]
	\centering
	\includegraphics[width=\textwidth]{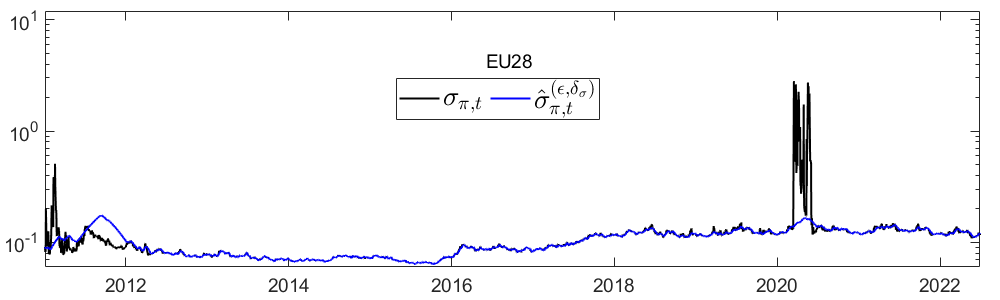}
	\includegraphics[width=\textwidth]{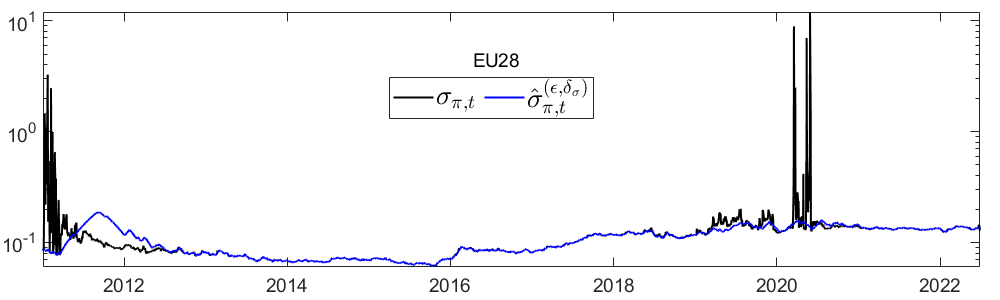}
	\caption{The EU28 time series $\sigma_{\pi,t}$ and $\hat{\sigma}_{\pi,t}^{(\epsilon,\delta_\sigma)}$
		for $\sigma_{jk}$ obtained via (top) \eqref{eq:sig_jk} and (bottom) \eqref{eq:reg_sig_jk}.}
	\label{fig:v9v11_sigma_EU}	
\end{figure}

Fig.~\ref{fig:v9v11_sigma_EU} compares the unregularized time series $\sigma_{\pi,t}$ from \eqref{eq:sigma_pi}
with the regularized time series
$\hat{\sigma}_{\pi,t}^{(\epsilon,\delta_\sigma)}
	= \sqrt{ \sum_{k=1}^{N-1} \left( \hat{\sigma}_{\pi k,t}^{(\epsilon,\delta_\sigma)} \right )^2 }$.
Shown are comparisons using the solutions \eqref{eq:sig_jk} and \eqref{eq:reg_sig_jk} for the matrix $\Sigma$.

\section{Results: Empirical Application}\label{sec:empirical}                                            

A projected use of the SRR is as a discriminator between asset classes,
with the implication that an asset class with higher values of $\nu_t$ is preferred.
We tested the estimated SRR for four different empirical datasets of stocks.
We also included a dataset of exchange-traded funds (ETFs).

\subsection{Variation with Asset Type}\label{sec:type}
Each dataset consisted of $N = 28$ assets.
The first asset group (denoted DJ28) is composed of the stocks in the Dow Jones Industrial Average
(excluding the  Dow Chemical Company and Visa Inc.,
for which the historical data --  from 20 March, 2019 and 18 March, 2008, respectively --
was too limited).\footnote{
	Hence the rational for setting $N = 28$ .
} 
The second (denoted SP28), third (denoted RS28) and fourth (denoted EU28) asset groups are composed of
a subset of the Standard and Poor's 500 Index, the Russell 3000 Index, and the Stoxx Europe 600 Index, respectively.
The last group (denoted ETF28) uses a selection of ETFs on US stocks.
To avoid selection bias, we utilize a consistent method to choose the subset of securities for each of SP28, RS28, EU28 and  ETF28.
Specifically we
\begin{itemize}
	\item ordered the assets in the universe in question by their market capitalization;
	\item parsimoniously, and as symmetrically as possible,
		removed the assets with the lowest and highest capitalization
		until the remaining number is divisible by 28; and
	\item picked the $(k-0.5)/28$, $k = 1, ..., 28$, percentile assets.
\end{itemize}

Each dataset consists of daily asset prices from 3 Jan., 2001 through 29 June, 2022.
We used  an estimation window of ten years (specifically 2500 trading days). 
Daily SRRs were computed from 3 Jan., 2011 through 29 June, 2022.
As noted in Section~\ref{sec:model}, we utilized the regularization parameter values
$\epsilon = 0.005$ and $\delta_\nu = 10^{-5}$.

\begin{figure}[htb]
	\centering
	\includegraphics[width=\textwidth]{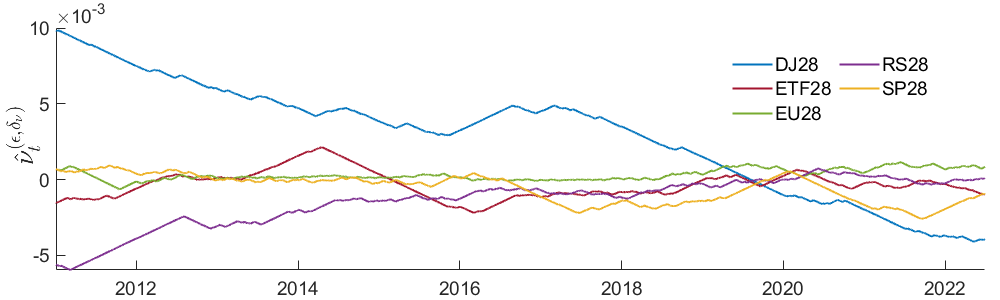}
		\caption{The time series  $\hat{\nu}_t^{(\epsilon,\delta_\nu)}$ for the five 28-asset groups.}
	\label{fig:srr_eps_del_28}	
\end{figure}
\begin{figure}[htb]
	\centering
	\includegraphics[width=0.49\textwidth]{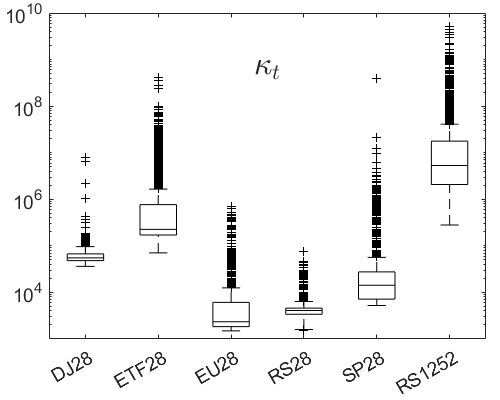}
	\includegraphics[width=0.49\textwidth]{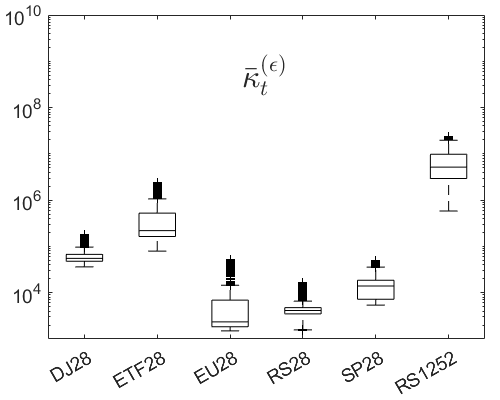}
	\includegraphics[width=0.49\textwidth]{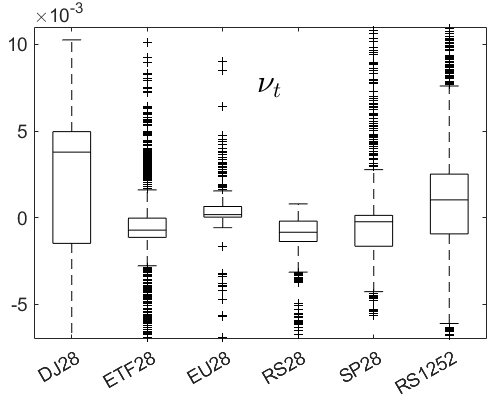}
	\includegraphics[width=0.49\textwidth]{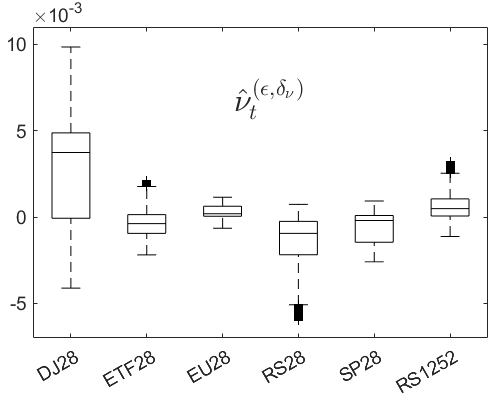}
		\caption{Box-whisker summaries of the unregularized and regularized time-series distributions
		of condition numbers and	SRR values for the five 28-asset groups and for the RS1252 group.
		The summaries for SRR are plotted using the y-axes  limits determined by the
		$\hat{\nu}_t^{(\epsilon,\delta_\nu)}$ box-whisker plots (bottom right),
		hence many outliers for $\nu_t$ are not shown in the bottom-left plot.
		}
	\label{fig:Ksrr_box_28_1252}	
\end{figure}
Fig.~\ref{fig:srr_eps_del_28} plots the regularized SRR time series $\hat{\nu}_t^{(\epsilon,\delta_\nu)}$
obtained for these five 28-asset groups.
Note that changes in the condition numbers of their respective matrices $\Phi_t$ over time
leads to a drift in the ``baseline'' behavior of $\nu_t$
(and, hence, for $\hat{\nu}_t^{(\epsilon,\delta_\nu)}$).
Fig.~\ref{fig:Ksrr_box_28_1252} presents box-whisker summaries of the distribution of values for
$\kappa_t$, $\bar{\kappa}_t^{(\epsilon)}$,  $\nu_t$, and $\hat{\nu}_t^{(\epsilon,\delta_\nu)}$ 
over this 11.5-year period.
The quantile values obtained for $\hat{\nu}_t^{(\epsilon,\delta_\nu)}$ are presented in Table~\ref{tab:q_nu_28_1252}.
The results strongly support the DJ28 group as the preferred investment group over 75\% of the 11.5-year period.
The EU28 group also has positive SRR values over 75\% of the 11.5-year period, but its daily SRR return rates
have a much narrower variation.
For the remaining three, the regularized SRR values are negative for more than 50\% of the 11.5-year time
period.
\begin{table}[htb]
	\caption{Quantile values of $\hat{\nu}_t^{(\epsilon,\delta_\nu)}$ from Fig.~\ref{fig:Ksrr_box_28_1252}.}
	\label{tab:q_nu_28_1252}
	\begin{tabular}{l cccccc}
	\toprule
	Quantile	& DJ28	& ETF27	& EU28	& RS28	& SP28 	& RS1252\\
	\midrule
	$P_{75}$	& $ 4.9 \cdot 10^{-3}$& $ 1.4 \cdot 10^{-4}$& $ 6.3 \cdot 10^{-4}$& $-2.5 \cdot 10^{-4}$& $ 9.3 \cdot 10^{-5}$ 
					& $ 1.1 \cdot 10^{-3}$ \\
	$P_{50}$	& $ 3.7 \cdot 10^{-3}$& $-3.8 \cdot 10^{-4}$& $ 1.9 \cdot 10^{-4}$& $-9.5 \cdot 10^{-4}$& $-2.0 \cdot 10^{-4}$
					& $ 4.9 \cdot 10^{-4}$ \\
	$P_{25}$	&$-7.1 \cdot 10^{-5}$& $-9.4 \cdot 10^{-4}$& $ 4.4 \cdot 10^{-5}$& $-2.2 \cdot 10^{-3}$& $-1.5 \cdot 10^{-3}$ 
					& $ 6.2 \cdot 10^{-5}$ \\
	\bottomrule
	\end{tabular}
\end{table}

\subsection{Dependence on Group Size}\label{sec:size}
We computed the SRR for a group of 1252 assets chosen from the Russell 3000 index.
Asset choice was determined by the requirement that asset prices cover the period 10 Jan., 2000
through 29 June, 2022.
Again, daily SRR values were computed from 3 Jan., 2011 through 29 June, 2022 using a moving window
of 2500 days.
Regularization was done using the same parameter values as in Section~\ref{sec:type} with
\eqref{eq:sig_jk} used for the computation of the elements of $\Sigma$.

\begin{figure}[htb]
	\centering
	\includegraphics[width=\textwidth]{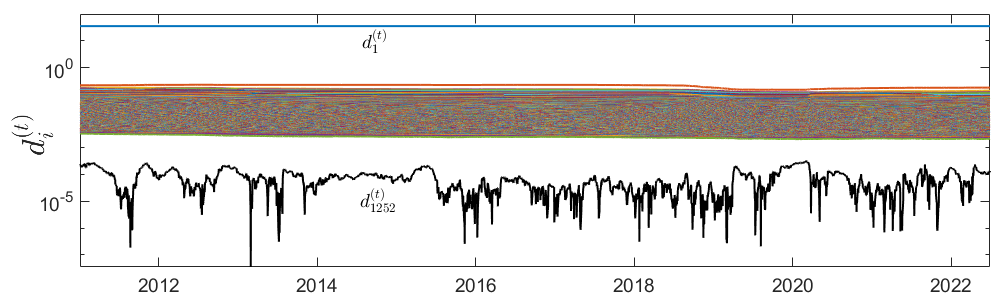}
	\caption{The time series of the singular values
		$d_j^{(t)}$, $j = 1, ..., 1252$ of the matrix $\Phi_t$
		 for the  1252-asset group.}
	\label{fig:dK_RS1252}	
\end{figure}
Fig.~\ref{fig:dK_RS1252} shows the time series of singular values $d_j^{(t)}$, $j = 1, ..., 1252$ of the matrix
$\Phi_t$ for this group.
As for the 28-asset groups discussed in Section~\ref{sec:type}, the behavior of the smallest singular value is responsible
for the behavior of the condition number of the 2891 $\times$ 1252 matrix $\Phi_t$.
From the behavior of $d_{1252}^{(t)}$, it is vairly obvious that the matrix $\Phi_t$ is both poorly conditioned
and has frequent, significant changes in condition number.

The respective plots in Fig.~\ref{fig:Ksrr_box_28_1252}  include the distribution of unregularized $\kappa_t$ and
regularized $\bar{\kappa}_t^{(\epsilon)}$ condition numbers seen over this time period for the 1252-asset group.
While for the 28-asset groups, the regularized condition number
$\bar{\kappa}_t^{\epsilon} \sim O(10^3) - O(10^5)$,
for the 1252-asset group, the regularized condition number is
$\bar{\kappa}_t^{\epsilon} \sim O(10^7)$ .
The respective plots in Fig.~\ref{fig:Ksrr_box_28_1252} also include the distributions of $\nu_t$ and
$\hat{\nu}_t^{(\epsilon,\delta_\nu)}$ for the RS1252 group.
In spite of its poorly behaved condition number, the regularized results for RS1252 show
improved regularized SRR values compared to the RS28 group.
The values of the quartiles of this distribution are also provided in Table~\ref{tab:q_nu_28_1252}.

\subsection{Behavior of the State-Price Deflator}\label{sec:spd}
As noted in Section~\ref{sec:model}, computation of the drift $\mu_{\pi,t}$ and total volatility $\sigma_{\pi,t}$
time series offers critical insight into the trajectory of the state-price deflator $\pi_t$.
As the SRR $\nu_t = - \mu_{\pi,t}$, we prefer to consider the behavior of the pair $\nu_t$ and $\sigma_{\pi,t}$,
using the regularized values $\hat{\nu}_t^{(\epsilon,\delta_\nu)}$ and $\hat{\sigma}_{\pi,t}^{(\epsilon,\delta_\sigma)}$.

\begin{figure}[htb]
	\centering
	\includegraphics[width=0.49\textwidth]{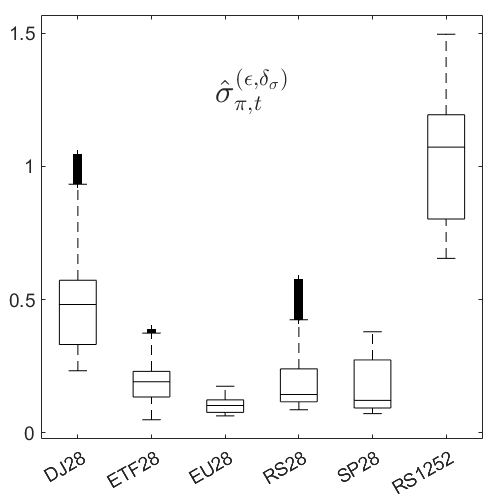}
	\includegraphics[width=0.49\textwidth]{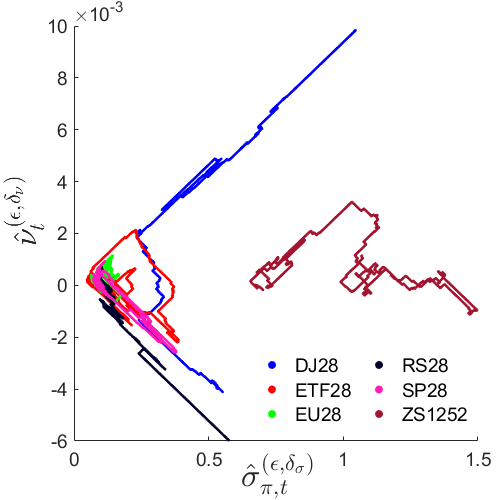}
	\caption{(left) Box-whisker summaries of the regularized time-series distributions of
		SS values for the five 28-asset groups and the RS1252 group.
		(right) The trajectories of the point
		$\left( \hat{\sigma}_{\pi,t}^{(\epsilon,\delta_\sigma)}, \hat{\nu}_t^{(\epsilon,\delta_\nu)} \right)$
		for the 1252-asset and 28-asset groups
	}
	\label{fig:ss_28_1252}	
\end{figure}

Comparisons of the distributions of the SRR values $\hat{\nu}_t^{(\epsilon,\delta_\nu)}$ for the six asset groups
analyzed in Sections~\ref{sec:type} and \ref{sec:size} were presented in Fig.~\ref{fig:Ksrr_box_28_1252}.
The distributions of the values $\hat{\sigma}_{\pi,t}^{(\epsilon,\delta_\sigma)}$ for the six asset groups
are presented in Fig.~\ref{fig:ss_28_1252}.
The volatility of the 1252-asset group is much larger than that of the 28-asset groups;
undoubtedly reflecting the fact that the deflator now has to ensure $\mathbb{P}-$martingale behavior of
a larger set of correlated assets.
Fig.~\ref{fig:ss_28_1252} displays the trajectories of the point
$\left( \hat{\sigma}_{\pi,t}^{(\epsilon,\delta)},  \hat{\nu}_t^{(\epsilon,\delta_\nu)} \right)$
over this time period for all six asset groups.
While there are time periods of linear correlation between  $\hat{\nu}_t^{(\epsilon,\delta_\nu)}$ and
$\hat{\sigma}_{\pi,t}^{(\epsilon,\delta)}$ (with either positive or negative slope) there are clear
``breakpoints'' when the behavior abruptly changes.

\begin{figure}[htb]
	\centering
	\includegraphics[width=\textwidth]{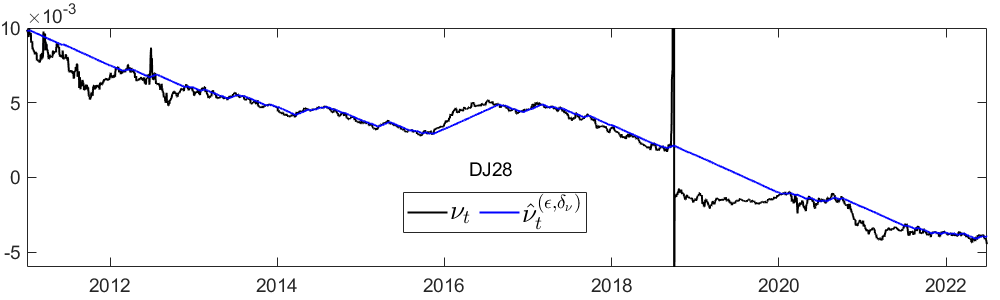}
	\includegraphics[width=\textwidth]{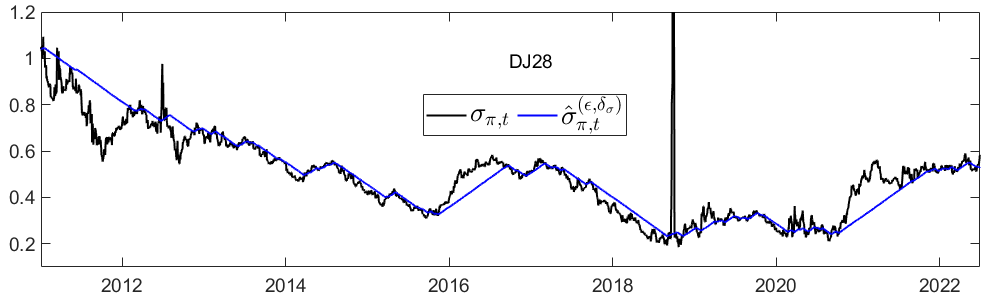}
	\caption{Time series for the regularized and unregularized values of $\nu_t$ and $\sigma_{\pi,t}$
		for the DJ28-asset group.
	}
	\label{fig:nu_sig_DJ28}	
\end{figure}
As the DJ28 trajectory is the simplest, we explore it in more detail.
Fig.~\ref{fig:nu_sig_DJ28} plots unregularized and regularized time series of $\nu_t$ and $\sigma_{\pi,t}$
for this 11.5 year time period.
With the exception of  the prominent spike over the period 24 Sept. to 1 Oct., 2018 (discussed in Section~\ref{sec:timeseries}),
the regularized solution smooths the overall downward trend in $\nu_t$,
while capturing the prominent ``bump'' that occurs over the period 30 Oct., 2015 through 21 Sept., 2018.
While the spike results in a lowering of the base value of $\nu_t$,
it correlates with a change from overall downward to flat-to-increasing behavior in $\sigma_{\pi,t}$.
There is also a prominent upward ``bump'' in $\sigma_{\pi,t}$ values over the period 30 Oct., 2015 through 21 Sept., 2018.

\begin{figure}[htb]
	\centering
	\includegraphics[width=0.49\textwidth]{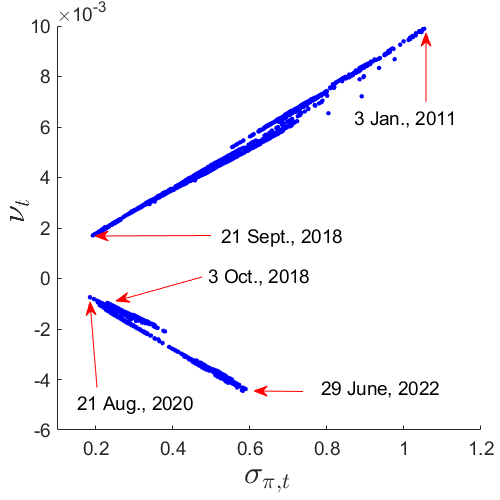}
	\includegraphics[width=0.49\textwidth]{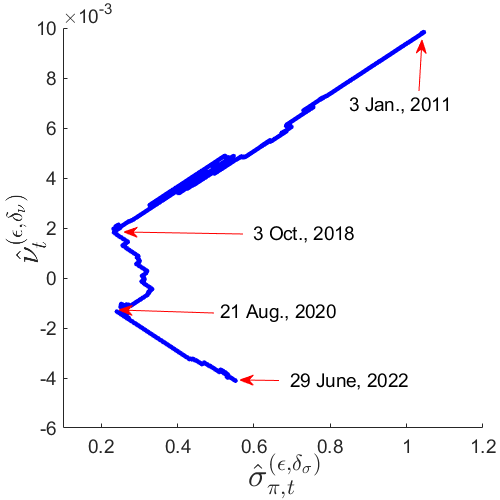}
	\caption{Trajectory of (left) $\left( \sigma_{\pi,t}, \nu_t \right)$ and (right)
		$\left( \hat{\sigma}_{\pi,t}^{(\epsilon,\delta_\sigma)}, \hat{\nu}_t^{(\epsilon,\delta_\nu)} \right)$
		for the DJ28-asset group, with significant time points indicated.
	}
	\label{fig:scatter_DJ28}	
\end{figure}
Fig.~\ref{fig:scatter_DJ28} summarizes the trajectories of
the unregularized time values $\left( \sigma_{\pi,t}, \nu_t \right)$ and the regularized time values
$\left( \hat{\sigma}_{\pi,t}^{(\epsilon,\delta_\sigma)}, \hat{\nu}_t^{(\epsilon,\delta_\nu)} \right)$
for the DJ28 group.
The range of axes values are determined by the regularized parameter plot on the right,
as a consequence seven outlier values for the unregularized parameter plot on the left are not seen.
These outlier values correspond to the spike values for the trading days 24 Sept. through 2 Oct., 2018.
The dates corresponding to the start and end of the spike discontinuity are shown in the unregularized
parameter plot.
The regularized parameter plot removes the discontinuity and provides a relaxation period
(3 Oct., 2018 through 21 Aug., 2020) to smooth the discontinuity.
Outside of this period, the regularized parameter trajectory provides a reasonably accurate smoothing of the
unregularized trajectory.

\section{Discussion} \label{sec:disc}
Our principal concern in this work has been the development of a numerical model to empirically compute
the shadow riskless rate described in Section~\ref{sec:model} for any selected universe of assets.
There are two significant challenges to this.
The first is the approximation of the volatilities $\sigma_{jk}$ in \eqref{eq:GBM} describing the stochastic effects
of the $N-1$ Brownian motions on each risky asset.
The method for estimating these parameters is detailed in Section~\ref{sec:calibration}.
The second is to control the ill-conditioned nature of the matrix $\Phi$ of the linear system \eqref{eq:linsys}
that uniquely determines the shadow riskless rate.
As a rough rule-of-thumb, a condition number of $\kappa(\Phi) = 10^k$  implies a potential loss of $k$ digits
of numerical accuracy in addition to the loss of precision expected by the LU decomposition (with pivoting)
used to solve \eqref{eq:linsys}.
There is, of course, no way to replace the linear system \eqref{eq:linsys} by a well conditioned linear system
that has the same solution set $\{x^{(i)}, i = 1, ... \}$ for any given input set $\{\mu^{(i)}, i = 1, ...\}$.
Rather, one has to develop an approximate (regularized) solution which controls changes in the
condition number of the system.
Our method for doing so, and its advantages, were presented in Sections~\ref{sec:reg} and \ref{sec:re_reg}.

A side benefit of the empirical work is that it provides insight into the time development of the drift and total
volatility of the state-price deflator for the asset group considered.
We have investigated this in Section~\ref{sec:spd}.

Given the relatively brief exploration of datasets in Sections~\ref{sec:type} and \ref{sec:size},
there are two important avenues requiring deeper investigation.
The first involves investigating the SRR for other asset classes (fixed income, commodities, crypto-assets, foreign currency, etc.)
as well as for a variety of geopolitical financial markets (China, United Kingdom, Japan, India, Hong Kong, etc.).
The second requires a more comprehensive investigation of the scaling of the SRR with the size of the asset group.
Of further-reaching importance to the asset-class discriminatory use of the SRR is the question of how well future values
(and how far into the future such values) can be predicted.
Such an approach might be accomplished by fitting a historical SRR time series ( appropriately modified to be stationary)
to an ARMA-GARCH model, and developing multi-day, out-of-sample, forecasts. 
Value-at-risk backtesting can then be applied to the out-of-sample projections to determine the loss of predictive
accuracy with increasing time of forecast.
As the assumption of asset prices governed by geometric Brownian motion is in fundamental disagreement with
observed price return distributions,
as in the theoretical work by \cite{Rachev_2017}, it will ultimately be necessary to develop a procedure to compute
SRR values  for risky assets following more realistic price processes, such as those described by i) jump diffusions,
ii) diffusions with local volatilty, and iii) geometric fractional Brownian or Rosenblatt motion.

Prior to pursuing these investigations, our work suggests a second avenue for defining a shadow riskless rate,
whose numerical solution would be much better conditioned.
We start with the observation that the column eigenvectors $w_{\circ j}$, $j = 1, ..., N$, obtained from the PCA analysis
of the historical return data for the $N$ underlying risky assets, are normalized.
Column vector $w_{\circ j}$ can be interpreted as a unique (and orthogonal) set of $N$ positions.
Application of these positions to the $N$ underlying risky assets
produces a composite asset ${\cal R}^{(j)}$ having return $r^{(j)} = \sum_{i=1}^N w_{ij} r_i$, $i = 1, ..., N$,
where $r_i$ is a historical return for underlying asset $i$.
Assume $N$ is large and let $1 < K \ll N$.

Consider the following algorithm for estimating a low-variance rate.

\smallskip\noindent
Find the minimum mean-variance portfolio $\sigma_p^{(K)}, r_p^{(K)}$ for the composite assets ${\cal R}^{(1)}, ..., {\cal R}^{(K)}$\\
$j = K$\\
continue $=$ true\\
while( continue )\\
\indent
	$j = j+1$\\
\indent
	Find the minimum mean-variance portfolio $\sigma_p^{(j)}, r_p^{(j)}$ for the composite assets ${\cal R}^{(1)}, ..., {\cal R}^{(j)}$\\
\indent
	if(  ($\sigma_p^{(j)} - \sigma_p^{(j-1)} > \textrm{tol}_\sigma$) or ($r_p^{(j)} - r_p^{(j-1)} > \textrm{tol}_r$ ) ) continue $=$ false\\
\noindent
end\\
$r = r_p^{(j)}$\\
$\sigma_r = \sigma_p^{(j)}$

\smallskip\noindent
This final minimum mean-variance portfolio provides a set of weights $q_r = \{ q_1^{(j)}, ..., q_j^{(j)}\}$ satisfying
$\sum_{i=1}^j q_i^{(j)} = 1$.\footnote{
	We assume long-only mean-variance optimization.
}

As the eigenvectors $w_{\circ j}$ are ordered according to the amount of variance of the data each explains,
successfully adding composite assets ${\cal R}^{(j)}$, $j = K, K+1, ...$ should slowly decrease both $\sigma_p^{(j)}$ and $r_p^{(j)}$
until a value of $j$ is reached such that successive additions merely add ``noise'' that will increase either
$\sigma_p^{(j)}$ or $r_p^{(j)}$.
The value $r$ is the minimum return possible (under the variance risk measure) for this asset class,
having minimum standard deviation $\sigma_r$.
We propose that $r$ can provide an empirical proxy to a minimum-risk rate for the class of assets under consideration.
The weights $q_r$ applied to the composite assets ${\cal R}^{(1)}, ..., {\cal R}^{(j)}$ define the minimum-variance
asset whose drift component is given by $r$.

In contrast, let $\left( r^{(N)}, \sigma_r^{(N)} \right)$ denote the minimum mean-variance portfolio achieved
by considering the full set $N$ of underlying assets each having historical return $r_i$.
The value of $\sigma_r$ should be smaller than that of $\sigma_r^{(N)}$ -- the question being ``can it be significantly smaller?''

\begin{appendices}

\renewcommand{\theequation}{A.\arabic{equation}}% reset appendix equation label
\setcounter{equation}{0}% reset the equation counter

\section{Principal Component Analysis}%[\appendixname~\thesection]{Principal Component Analysis}
Consider the data matrix $X \in \mathbb{R}^{M \times N}$,
where each row vector $x_{m\circ}$ of $X$ represents the values obtained from a sample $m$, $m = 1, ..., M$,
each sample of size $N$.
We assume each column vector $x_{\circ j}$ of $X$ satisfies $\frac{1}{M} \sum_{m=1}^M x_{mj} = 0$
(i.e., the sample mean of each column has been shifted to zero).
Let column $w_{\circ j}$, $j = 1, ..., N$, of the matrix $W \in \mathbb{R}^{N \times N}$ denote the orthonormal
eigenvectors of $X^{\text{T}} X$.
Then $X^{\text{T}} X W = W \Lambda$, where $\Lambda = \text{diag}(\lambda_1, ..., \lambda_N)$
contains the eigenvalues.
The eigenvalue-eigenvector pairs are ordered such that $\lambda_1 \ge \lambda_2 \ge ... \ge \lambda_N \ge 0$.\footnote{
	The matrix $X^{\text{T}} X$ is positive semidefinite.
}
The matrix $P \in \mathbb{R}^{M \times N}$ satisfying $P = X W$ has the property that
$p_{mj}$ represents the projection of row vector $x_{m\circ}$ on eigenvector $w_{\circ j}$.
Thus the components $p_{mj}$, $m = 1, ..., M$, of the column vector $p_{\circ j}$ of $P$ satisying $p_{\circ j} = X w_{\circ j}$ represent the
projections of the vectors $x_{m \circ}$ (i.e., the results of experiment $m$) on the eigenvector $w_{\circ j}$.
The vectors $p_{\circ j} = X w_{\circ j}$ are known as the principal components of the analysis.
The sample covariance between two different principal components is zero by the orthogonality of the eigenvectors,
\begin{equation}\label{eq:Qpp}
\text{Cov}(p_{\circ i}, p_{\circ j}) = (X w_{\circ i})^{\text{T}} (X w_{\circ j})
	= \lambda_j  w_{\circ i}^{\text{T}} w_{\circ j} = \lambda_j  \delta_{ij}.
\end{equation}
Thus the principal component covariance matrix $C_P$ is $P^{\text{T}}P = \Lambda$.
Principal component analysis is a transformation to coordinates $w_{\circ j}, j = 1, ..., N$,
which diagonalizes the empirical covariance matrix.

As $ X^{\text{T}} X W = W \Lambda = W \Lambda W^{\text{T}} W$,
where the orthonormal columns $w_{\circ j}$ of $W$ span $\mathbb{R}^N$,
we see that the covariance matrix $C_X$ of the empirical data is expressed as
\begin{equation}\label{eq:pca_cov}
C_X = X^{\text{T}} X = W \Lambda W^{\text{T}} = \left(W \sqrt{\Lambda} \right) \left( W \sqrt{\Lambda} \right)^{\text{T}}.
\end{equation}

Thus, for the general case $M > N$, principal component analysis identifies a reduced space $\mathbb{R}^{N \times N}$
spanned by an orthonormal set of vectors $w_{\circ j}$, $j = 1, ..., N$.
The spanning vectors $w_{\circ j}$ are eigenvectors of the data matrix $X^{\text{T}} X$,
ordered such that the first principal component $p_{\circ 1} = X w_{\circ 1}$ explains the greatest
variance of the data, the second principal component $p_{\circ 2} = X w_{\circ 2}$ explains the next largest amount of variance, etc.
\end{appendices}

\bibliographystyle{apacite} %{acm}
\bibliography{SRR}

\end{document}